\documentclass[useAMS,usenatbib,usegraphicx]{mn2e}

\pagerange{\pageref{firstpage}--\pageref{lastpage}}
\pubyear{2009}

\newcommand{\be}{\begin{equation}}
\newcommand{\ee}{\end{equation}}

\begin{document}


\title[New estimates of the CMB angular power spectra from the WMAP 5 yrs low resolution data]
{New estimates of the CMB angular power spectra from the WMAP 5 yrs low resolution data}

\author[A.Gruppuso, A.De Rosa, P.Cabella, F.Paci, F.Finelli, P.Natoli, G.de Gasperis and N.Mandolesi]
{A.~Gruppuso $^{1,2}$\thanks{E-mail: gruppuso@iasfbo.inaf.it}, A.~De Rosa $^{1}$, P.~Cabella $^{3,4}$, F.~Paci $^{5,1,2}$ , 
F.~Finelli $^{1,7,2}$, 
\newauthor P.~Natoli $^{3,6}$, G.~de Gasperis $^{3}$ and N.~Mandolesi $^{1}$ 
\\
$^1$ 
INAF-IASF Bologna, Istituto di Astrofisica Spaziale e Fisica Cosmica 
di Bologna \\
Istituto Nazionale di Astrofisica, via Gobetti 101, I-40129 Bologna, Italy \\
$^2$
INFN, Sezione di Bologna,
Via Irnerio 46, I-40126 Bologna, Italy \\ 
$^3$
Dipartimento di Fisica, Universit\`a di Roma ÒTor VergataÓ, Via della Ricerca Scientifica 1, 00133 Roma, Italy \\
$^4$ 
Dipartimento di Fisica, Universit\`a di Roma ÒLa SapienzaÓ, P.le Aldo Moro 2, 00185 Roma, Italy \\
$^5$
Dipartimento di Astronomia, Universit\`a degli Studi di Bologna,
via Ranzani, 1 Ð I-40127 Bologna, Italy \\
$^6$
INFN, Sezione di Roma ÒTor VergataÓ, Via della Ricerca Scientifica 1, 00133 Roma, Italy \\
$^7$ INAF-OAB, Osservatorio Astronomico di Bologna
Istituto Nazionale di Astrofisica, via Ranzani 1, I-40127 Bologna, Italy \\
}

\label{firstpage}

\maketitle

\begin{abstract}
A Quadratic Maximum Likelihood (QML) estimator is applied to the WMAP 5 
year low resolution maps to
compute the CMB angular power spectra at large scales for both temperature 
and polarization. 
Estimates and error bars for the six angular power spectra are provided 
up to $\ell=32$
and compared,
when possible, to those obtained by the WMAP team, without finding any inconsistency.
The conditional likelihood slices are also computed for 
the $C_{\ell}$ of all the six power spectra from $\ell =2$ to $10$
through 
a pixel based likelihood code.
Both 
the codes
treat the covariance for $(T,Q,U)$ in a single matrix without employing any approximation.
The inputs of both the codes (foreground reduced maps, related covariances and masks) are provided by
the WMAP team.
The peaks of the likelihood slices are always consistent with the 
QML estimates within the error bars, however an excellent agreement occurs when 
the QML estimates are used as fiducial power spectrum 
instead of the best-fit theoretical power spectrum. 
By the full computation of the conditional likelihood on 
the estimated spectra, the value of the temperature quadrupole $C_{\ell=2}^{TT}$ is found to be less than $2 \, \sigma$ away 
from the WMAP 5 yrs $\Lambda$CDM best-fit value. 
The $BB$ spectrum is found well consistent with zero and upper limits on the B-modes are provided.
The parity odd signals $TB$ and $EB$ are found consistent with zero. 
\end{abstract}


\begin{keywords}
cosmic microwave background - cosmology: theory - methods: numerical - methods:
statistical - cosmology: observations
\end{keywords}

\section{Introduction}
The Cosmic Microwave Background (CMB hereafter)  is a powerful tool for
investigating the properties of the early and present Universe.
Under hypothesis of Gaussianity and statistical isotropy, the main cosmological
information coming from a CMB map is contained in its temperature and polarization
angular power spectra (APS hereafter).
In recent years several experiments, with different detection techniques, have
measured the CMB anisotropies with increasingly better 
precision (\cite{hinshaw1, kogut1, Bo, Bo1, Bo2, CBI,
MAXIPOL,  ACBAR, QUAD, QUAD1, QUAD2, dunkley_wmap5, page, hinshaw}).

Many methods have been developed to give unbiased estimates of the CMB power spectra
{\bf as} Master \citep{master}, Cross-Spectra (\cite{saha}, \cite{polenta}, \cite{grain}). 
At high multipoles ($\ell > 30$, \cite{efstathiou}) the so called $pseudo-C_l$ algorithms are usually preferred to others techniques.
These methods in fact implement
the estimation of
power spectral densities from periodograms \citep{hauser}.
Basically, 
they
estimate the $C_l$ through the inverse Harmonical transform
of a masked map that is then deconvolved with geometrical kernels and corrected with a noise bias
term.
These techniques
give unbiased estimates, and it has been shown they work
successfully when applied to real data at high multipoles \citep{Bo,MAXIPOL,ACBAR, dunkley_wmap5}. 
However, it is well known that at low multipoles they are not optimal since they provide power spectra 
estimates with error bars larger than the minimum variance.
Several 
strategies
for measuring $C_{\ell}$ at low resolution have been developed and
applied to CMB data with excellent results. These methods include the Quadratic Maximum Likelihood (QML)
estimator \citep{tegmark_tt, tegmark_pol}, and different sampling techniques such as Gibbs \citep{jewell, wandelt, eriksen}, 
adaptive importance \citep{teasing} and Hamiltonian \citep{taylor}. 

In this paper, 
we apply a parallel implementation of the QML estimator \citep{tegmark_tt,tegmark_pol} on WMAP 5 year data.
It is shown in \citep{efstathiou} that at low multipoles, the 
intrinsic variance introduced by this technique is lower than the
variance introduced by the the pseudo-$C_{\ell}$ methods. 
We also compute the conditional likelihood slices of the $C_{\ell}$ 
multipoles and compare the results with those obtained by the QML.

The largest angular scales of the CMB anisotropies probe the Physics of the early Universe.
Therefore estimating its power spectra with optimal methods is crucial. 
The temperature and polarization low multipole pattern contains information on the reionization process, dark energy and the geometry of the universe. Moreover, determining at best cosmological parameters as the optical depth $\tau$, has an important impact on the amplitude and the spectrum of 
primordial perturbations, and therefore on inflationary models. Optimal methods at low resolution are also required 
by the observed value of the temperature quadrupole, anomalously low 
within the $\Lambda$CDM dominant model which remains however the simplest cosmological model preferred by data.



In this paper we perform an
analysis of the power spectra from low resolution maps of 
the five year WMAP data providing $C_\ell$ and errors bars for all
the $6$ APS from $\ell=2$ to $32$ computed by our QML and we show the robustness of our results with respect to iterative
estimates. 
We also give the conditional likelihood slices computed with 
a pixel based likelihood code from $\ell=2 $ to $10$.
The paper is organized as follows: 
in Section \ref{sect:method} we describe the main equation of the QML and in Section \ref{sect:likelihood} we provide
the basic expression for the likelihood.
Section \ref{sect:comp} is dedicated to the technical implementation with a brief description of the computational requirements 
both for the QML and for the likelihood codes. In Section \ref{sect:data} we report the results obtained from WMAP 5 year data and
finally in Section \ref{sect:concl} we draw our conclusions.
As a convention, all the objects that are defined in pixel space are represented with bold face symbols.

\section{The QML method}
\label{sect:method}
The QML method for Power Spectrum Estimation (PSE) of CMB anisotropies was introduced in \citep{tegmark_tt} 
and was extended to polarization in \citep{tegmark_pol}. 
Given a map in temperature and polarization ${\bf x=(T,Q,U)}$, the QML provides estimates
$\hat {C}_\ell^X$ - with $X$ being one of $TT, EE, TE, BB,
TB, EB$ - of the APS as: 
\be
\hat{C}_\ell^X = \sum_{\ell' \,, X'} (F^{-1})^{X \, X'}_{\ell\ell'} \left[ {\bf x}^t
{\bf E}^{\ell'}_{X'ì} {\bf x}-tr({\bf N}{\bf
E}^{\ell'}_{X'}) \right]
\, ,
\ee
where the $F_{X X'}^{\ell \ell '}$ is the Fisher matrix defined as
\be
\label{eq:fisher}
F^{\ell\ell'}_{X X'}=\frac{1}{2}tr\Big[{\bf C}^{-1}\frac{\partial
{\bf C}}{\partial
  C_\ell^X}{\bf C}^{-1}\frac{\partial {\bf C}}{\partial
C_{\ell'}^{X'}}\Big] \,.
\ee
The ${\bf E}^{\ell}_X$ matrix is given by
\be
\label{eq:Elle}
{\bf E}^\ell_X=\frac{1}{2}{\bf C}^{-1}\frac{\partial {\bf C}}{\partial
  C_\ell^X}{\bf C}^{-1} \, ,
\ee
with ${\bf C} ={\bf S}(C_{\ell})+{\bf N}$ being the global covariance matrix (signal plus noise contribution, 
possibly extended to include residuals from foreground subtractions) and $C_\ell$ is called fiducial power 
spectrum. It can be shown that the QML method is unbiased in finding the APS $C_\ell^{X,obs}$ of the map ${\bf x}$:
\be
\langle\hat{C}_\ell^X\rangle=C_\ell^{X,obs} \,.
\label{unbiased}
\ee
It is also optimal,
since it can provide the
smallest error bars allowed by the Fisher-Cramer-Rao inequality,
\be
\langle\Delta\hat{C}_\ell^X
\Delta\hat{C}_{\ell'}^{X'} \rangle= ( F^{-1})^{X \, X'}_{\ell\ell'} \,  ,
\label{minimum}
\ee
where the averages are meant to be over an ensemble of realizations.

We have verified by Montecarlo on simulated data that our implementation of the QML method (called {\it BolPol}, see Section \ref{sect:comp}) leads to unbiased 
minimum variance results as from Eqs. (\ref{unbiased}, \ref{minimum}). We have also verified that 
Eq. (\ref{unbiased}) holds with $C_{\ell} \ne C_{\ell}^{obs}$, but the errors are no more minimum variance.


\section{Conditional Likelihood (or Probability) Slices}
\label{sect:likelihood}

Assuming that CMB anisotropies are Gaussian distributed, the likelihood for CMB temperature and polarization power spectra is given by
\cite{bond}:
\be
{\cal L} = exp \left[ -{1\over 2} {\bf x}^t {\bf C}^{-1} {\bf x}\right] / \sqrt{ (2 \pi )^n det ({\bf C})}
\label{Likelihood}
\, ,
\ee
where $n$ is the dimension of the vector ${\bf x}$ (and depends on the number of pixels that are observed).
It gives the probability to have the map ${\bf x}$ given the model (i.e. $C_{\ell}$) needed to build the signal covariance matrix.
The conditional likelihood slice of $C_{\bar \ell}^{\bar X}$ for given ${\bar \ell} \,, {\bar X}$ 
are obtained sampling Eq. (\ref{Likelihood}) at different values 
of $C_{\bar \ell}^{\bar X}$ keeping $C_{\ell}^{X}$ fixed for $\ell \ne {\bar \ell} \,, X \ne {\bar X}$.
Our implementation in parallel fortran 90 of eq.~(\ref{Likelihood}) is called {\it BoLike} (see Section \ref{sect:comp}).
It is important to stress that the covariance matrix in ${\bf x}$
either in {\it BolPol} or {\it BoLike} 
is treated without any approximation and not separated in ${\bf T}$ and $({\bf Q},{\bf U})$ as in \citep{page}.
BoLike can also be employed as part of a likelihood code, 
to compute directly the posterior probability of a cosmological 
model at low $\ell$ directly in pixel space without
relying on any approximation for the probability distributions 
of the $C_\ell^X$.

Other methods have been developed to compute the CMB likelihood for
low resolution dataset, which can also be used to provide power
spectrum estimates. One such approach is based on the Gibbs algorithm
\citep{wandelt,2005PhRvD..71j3002C}, which provides a
computationally feasible framework to sample the posterior probability
of the $C_\ell$ given the data, $P(C_\ell |\mathbf{x})$. Specifically,
this is achieved by repeatedly drawing samples from the (easier to
compute) conditional distributions $ P(\mathbf{s}|C_\ell,\mathbf{x})$
and $P(C_\ell,\mathbf{s})$, $\mathbf{s}$ being a signal only CMB map.
A Blackwell-Rao approximation can then be used to compute the CMB
likelihood \citep{2009ApJ...692.1669R}. This method has been applied to
the WMAP5 temperature data \citep{dunkley_wmap5} and the WMAP3
temperature plus polarization data \citep{2007ApJ...665L...1E}. One
feature of the Gibbs approach is that it can internally perform a
parametric based component separation, and propagate model
uncertaintes to power spectra and likelihoods
\citep{2007AAS...211.9001D}.

Another proposed approach to estimate the APS and likelihood at large
angular scales is based on adaptive importance sampling. 
The idea is to
model each single $C_\ell$ distribution with an inverse gamma
distribution, finding appropriate parameters. These distributions are
then used to provide an approximation to the global posterior. The
method so far has been demonstrated to work for intensity data and
applied to WMAP5 \citep{teasing}.



\section{Computational requirements}
\label{sect:comp}
{\it BolPol} is a fully parallel implementation of the QML method described in Section \ref{sect:method}, written in F90.
Since the method works in pixel space the computational cost increases as one considers smaller angular
resolution for a given sky area, i.e. more pixels. This is the reason why we parallelize the QML code. 
The inversion of the covariance matrix ${\bf C}$ scales as ${\cal O} (N_{\rm pix}^3)$.
The number of operations is roughly driven, once the inversion of the total covariance
matrix is done, by the matrix-matrix multiplications to build the operators ${\bf E}^{X}_{\ell}$ in 
Eq. (\ref{eq:Elle}) and by calculating 
the Fisher matrix $F_{X X'}^{\ell \ell'}$ given in eq. (\ref{eq:fisher}). 
The number of operations that are needed to build these matrices scales as $O( N_{side}^2 N^{2}_{pix})$ and $O( N_{side} N^{3}_{pix})$
respectively. 
The RAM required is of the order $O( \Delta \ell \, N^2_{pix})$ where $\Delta \ell$ is 
the number of ${\bf C}^{-1} \, (\partial {\bf C} / \partial C_{\ell}^X)$ (for every $X$)
that are built and kept in memory during the execution time. 
 

Given these kind of scalings, it is clear that it is unrealistic to run the QML estimator for all-sky maps of resolution larger than 
$N_{side}=8$ (in Healpix language\footnote{http://healpix.jpl.nasa.gov/. For the reader not familiar with the Healpix notation, $N_{side}$ is related to the number of pixels $N_{pix}$
by $N_{pix}= 12 N_{side}^2$.} \citep{gorski})
on a single CPU. 
To reach higher resolution we use the ScaLapack package and the BLACS routines.
In this way it is possible to run {\it BolPol} on the WMAP data set with the resolution of $N_{side}=16$ 
\footnote{Note that $N_{side}=16$ is not the highest resolution that {\it BolPol} is able to consider.
Currently {\it BolPol} is able to process maps of $N_{side}=32$.}
on a supercomputer like BCX (at CINECA, 
Bologna, processor type: Opteron Dual Core 2.6 GHz, with $4$ GB per processor) in $\sim 40$ minutes using 64 processors.

The computational resources required by the likelihood evaluation in {\it BoLike} are driven by the 
calculation and the inversion of total covariance matrix. 
{\it BoLike} shares with {\it BolPol} the routines for the computation and inversion of the covariance matrix.

\section{Description of the WMAP 5 years dataset and Results}
 \label{sect:data}

In this Section we describe the data set that we have considered. 
We use the ILC map in $T$, the foreground cleaned maps in $(Q,U)$, the 
noise covariance matrix in $(Q,U)$ and the masks
at $N_{side}=16$ publicly available at the LAMBDA web site
\footnote{http://lambda.gsfc.nasa.gov/}. The temperature map is the ILC map at $N_{side}=16$ smoothed at $9.8$ degrees.
We have added a random noise realization with variance of $1 \mu K^2$ as suggested in \cite{dunkley_wmap5}.
Consistently the noise covariance matrix for TT is taken to be diagonal with variance equal to $1 \mu K^2$.
We have explicitly checked that the estimates do not depend on the noise realization we added on the temperature map.
The temperature map has been masked with a mask covering $\sim 16 \, \%$ of the sky and the monopole and the dipole have been subtracted from
the observed sky by means of the Healpix routine ``remove\_dipole"  \citep{gorski} that works in pixel space.
The polarization maps for Q and U are provided by the WMAP team at the same resolution $N_{side}=16$.
The inverse of the masked noise covariance matrix for the polarization part is available at the same resolution. 
We have followed the procedure explained at the LAMBDA web site (see footnote 4)
to obtain
the direct noise covariance for the observed pixels. The used polarization mask is larger than the temperature mask
and covers $\sim 26 \, \%$ of the sky.
The noise covariance matrices for TQ and TU are not provided and we set them to zero.
In this work we use the WMAP 5 yrs publicly available data products to derive our spectral estimates. The uncertainties 
due to foreground cleaning cannot be taken into account since are not provided explicitly by the WMAP team.
The QML estimates have been obtained by constructing the signal covariance matrix by using fiducial 
$C_\ell$s up to $3 N_{\rm side}$ and we self-consistently compute $C_\ell^X$ up to $3 N_{\rm side}$. 
We show the QML estimates up to $2 N_{\rm side}$, which is a conservative choice for $\ell$-range, not to incur into
discretization errors.

\begin{figure*}
\includegraphics[width=15cm,height=10cm,angle=0]
{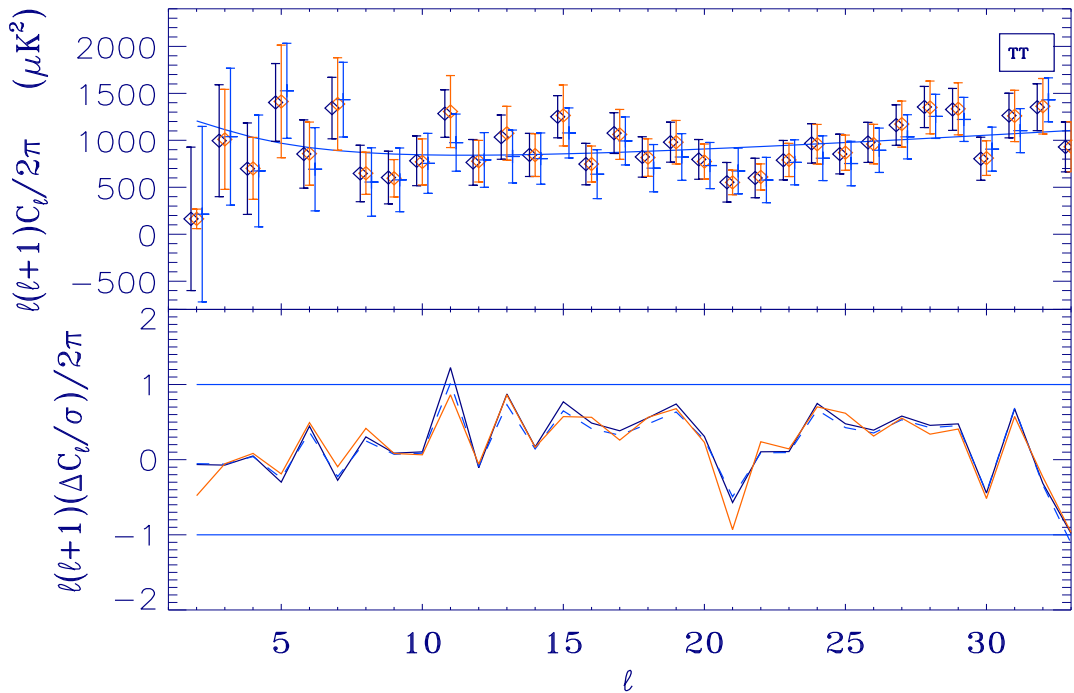}
\label{figPSestTT}
\caption{Estimates of TT angular power spectrum from WMAP 5 year data at low resolution.
Upper panel: {\it BolPol} estimates (dark blue diamonds) with error bars (dark blue), iterated {\it BolPol} estimates (red diamonds) with error bars (red),
WMAP pseudo-$C_\ell$ estimates (light blue cross) with error bars (light blue).
Lower panel: differences between the sets of estimates in unit of sigma (same conventions as upper panel for the colors).}
\end{figure*}

\begin{figure*}
\label{figPSestEE}
\includegraphics[width=15cm,height=10cm,angle=0]
{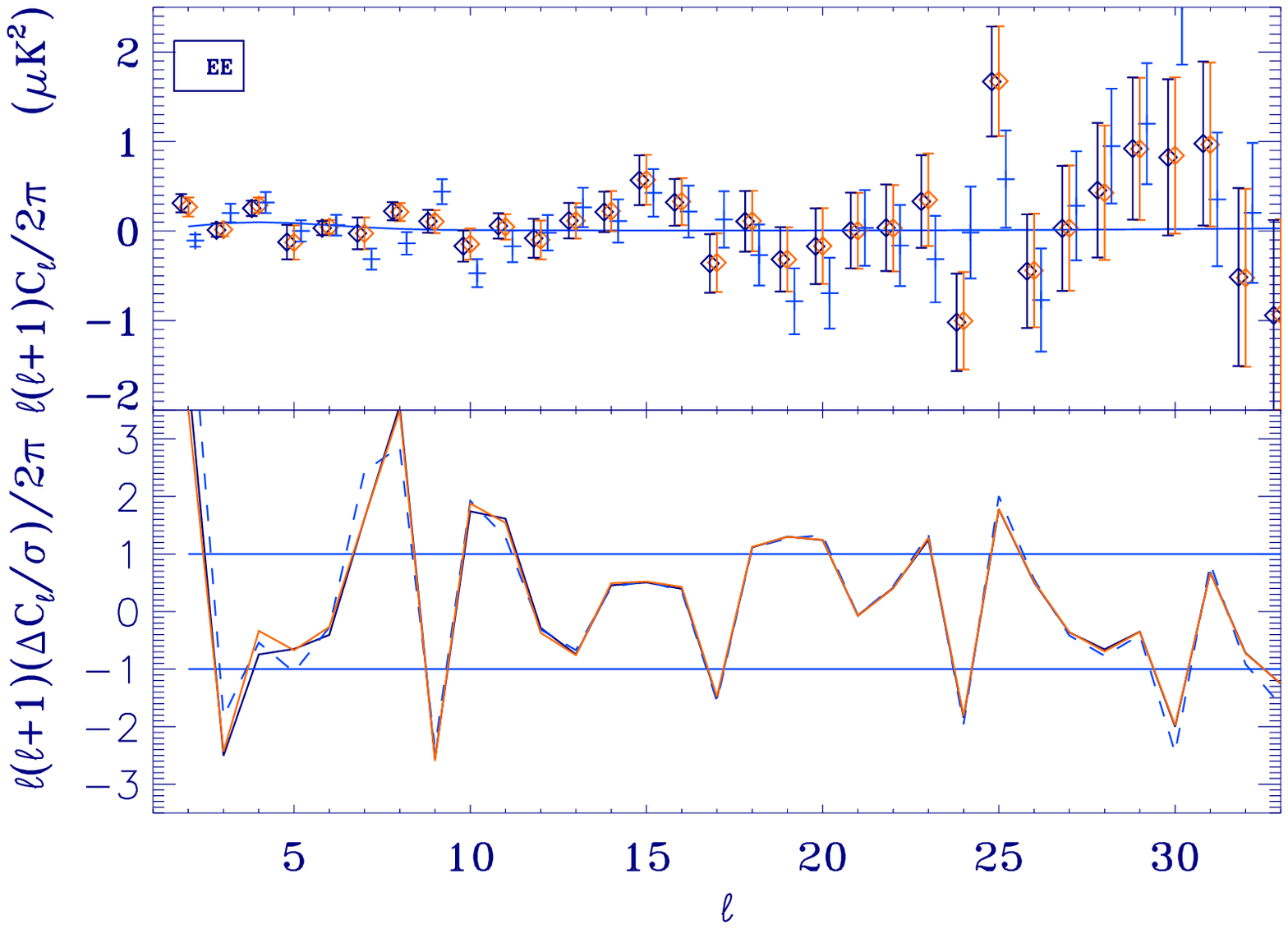}
\caption{Estimates of EE angular power spectrum from WMAP 5 year data at low resolution.
Upper panel: {\it BolPol} estimates (dark blue diamonds) with error bars (dark blue), iterated {\it BolPol} estimates (red diamonds) with error bars (red),
WMAP pseudo-$C_\ell$ estimates (light blue cross) with error bars (light blue).
Lower panel: differences between the sets of estimates in unit of sigma (same conventions as upper panel for the colors).}
\end{figure*}

\begin{figure*}
\label{figPSestTE}
\includegraphics[width=15cm,height=10cm,angle=0]
{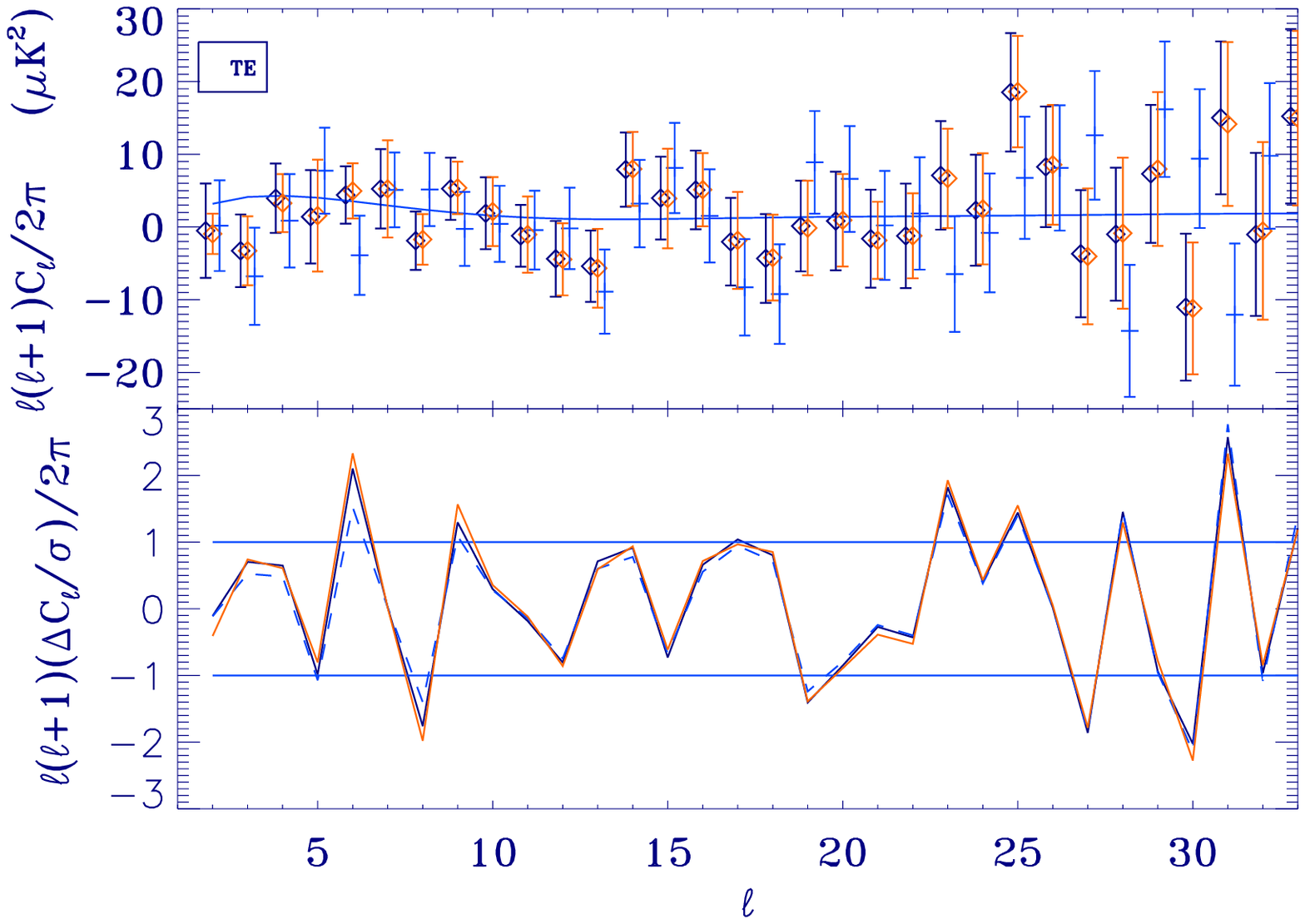}
\caption{Estimates of TE angular power spectrum from WMAP 5 year data at low resolution.
Upper panel: {\it BolPol} estimates (dark blue diamonds) with error bars (dark blue), iterated {\it BolPol} estimates (red diamonds) with error bars (red),
WMAP pseudo-$C_\ell$ estimates (light blue cross) with error bars (light blue).
Lower panel: differences between the sets of estimates in unit of sigma (same conventions as upper panel for the colors).}
\end{figure*}

\begin{figure*}
\label{figPSestBB}
\includegraphics[width=15cm,height=10cm,angle=0]
{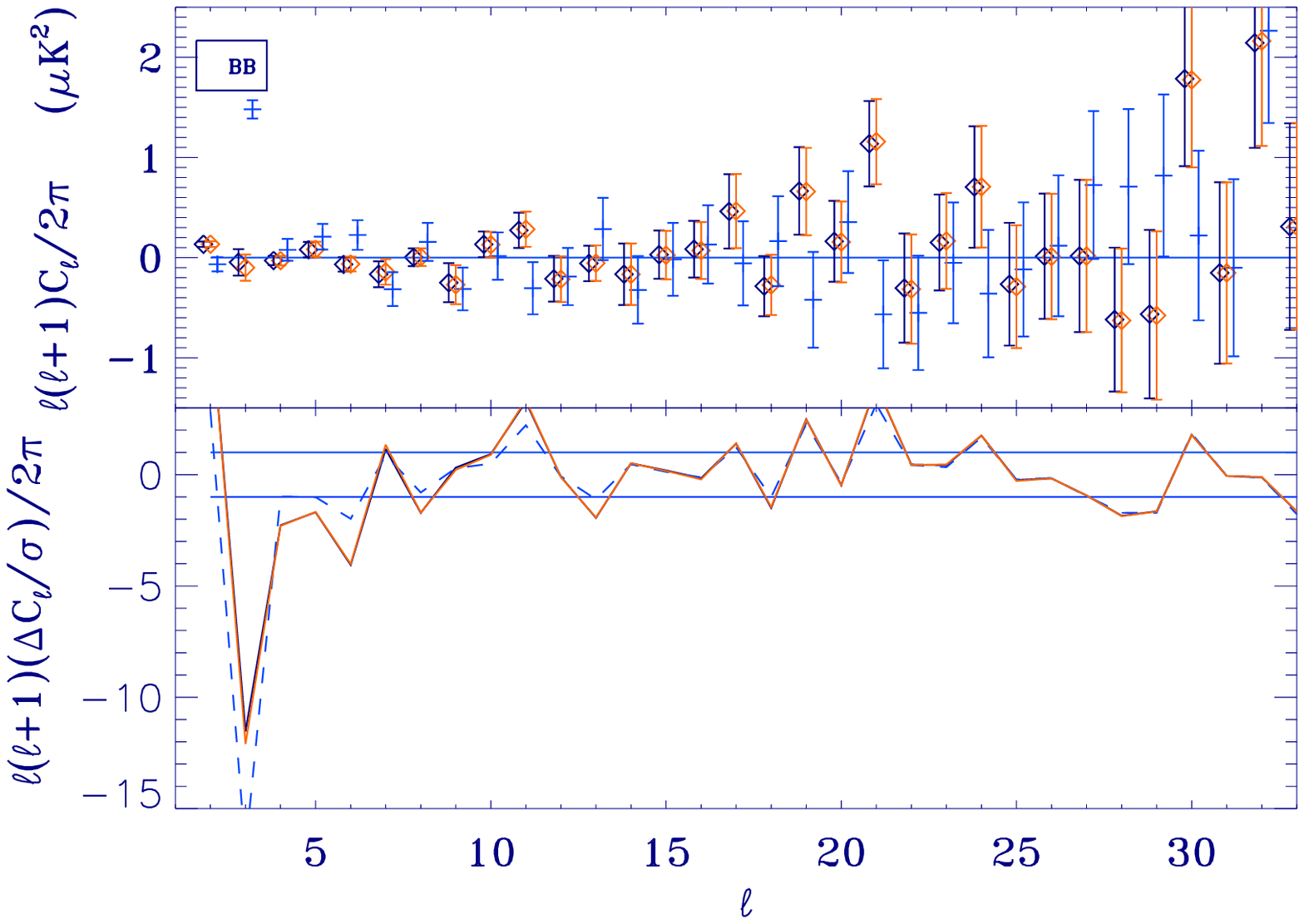}
\caption{Estimates of TT angular power spectrum from WMAP 5 year data at low resolution.
Upper panel: {\it BolPol} estimates (dark blue diamonds) with error bars (dark blue), iterated {\it BolPol} estimates (red diamonds) with error bars (red),
WMAP pseudo-$C_\ell$ estimates (light blue cross) with error bars (light blue).
Lower panel: differences between the sets of estimates in unit of sigma (same conventions as upper panel for the colors).}

\end{figure*}

\begin{figure*}
\label{figPSestTB}
\includegraphics[width=15cm,height=10cm,angle=0]
{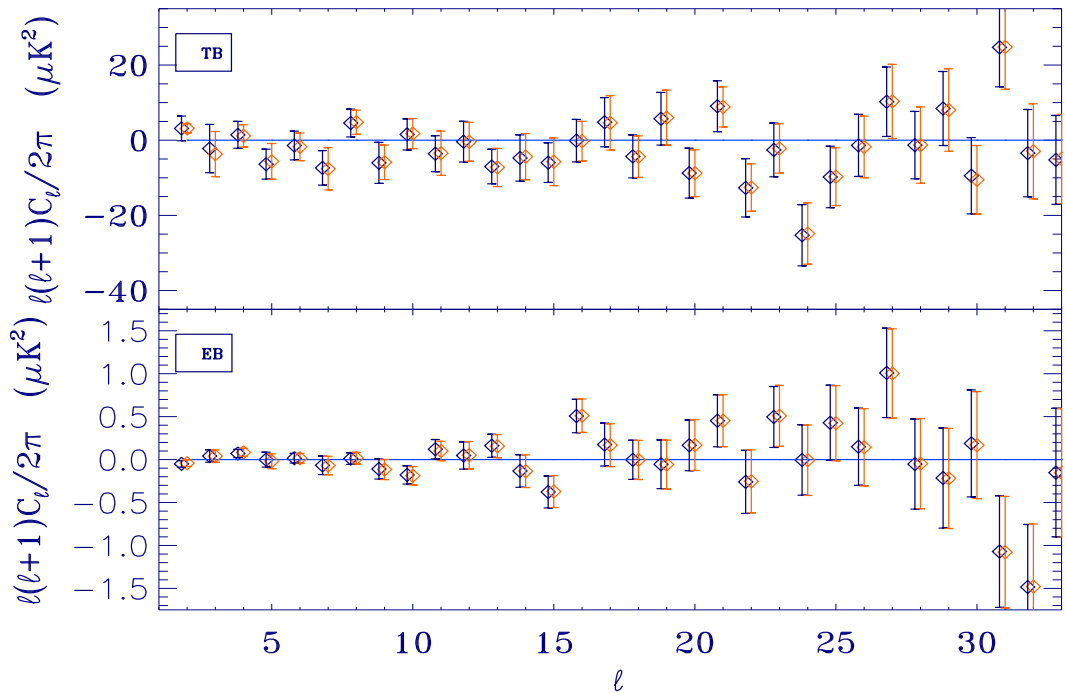}
\caption{{\it BolPol} estimates of TB (upper panel) and EB (lower panel) angular power spectra from WMAP 5 year data at low resolution.
Dark blue symbols are for the not iterated case and red for the iterated case.}
\end{figure*}

\begin{figure*}
\label{fig:slices1}
\includegraphics[width=18cm,height=15cm,angle=0]{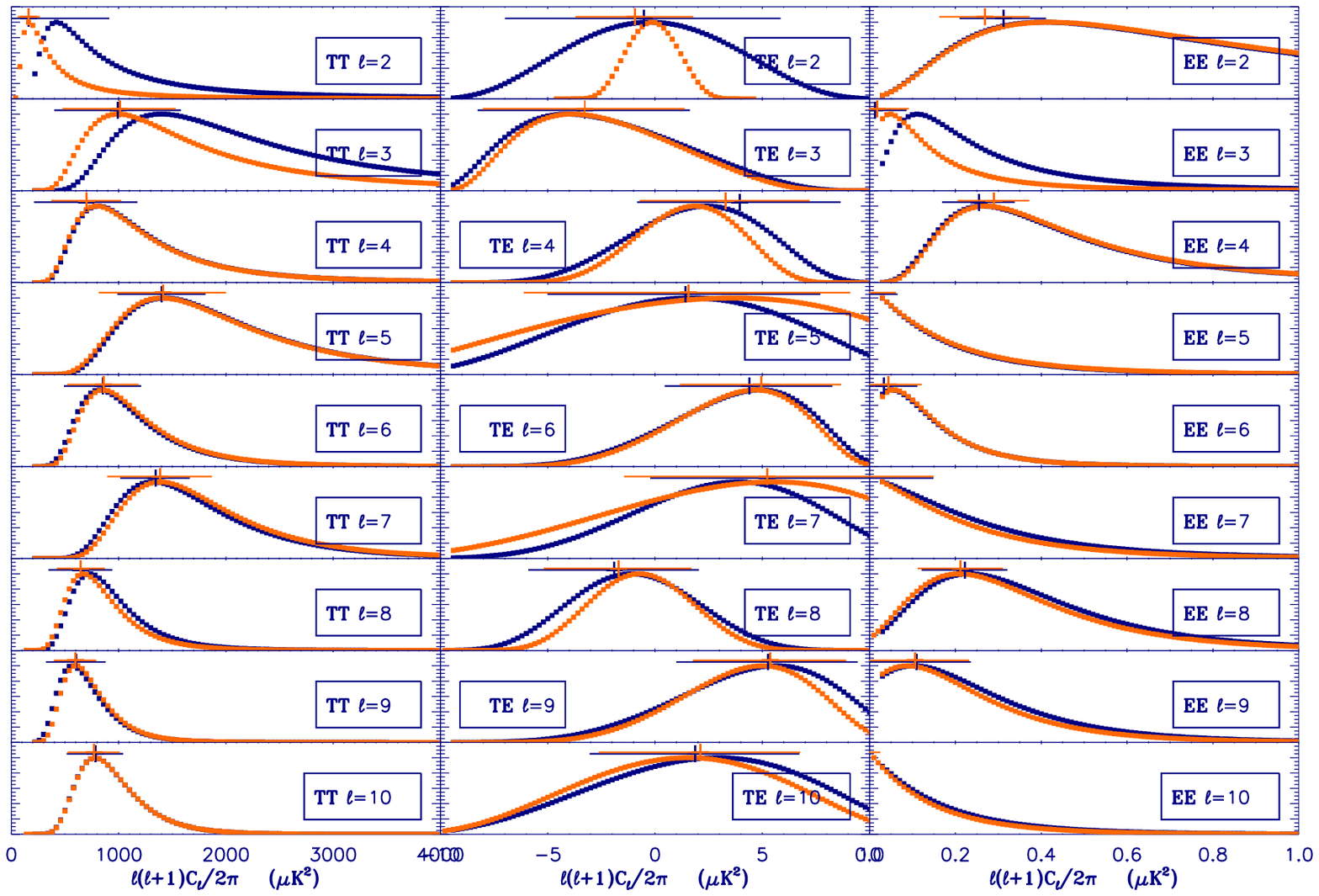}

\caption{Likelihood Slices for TT, TE and EE from $\ell=2$ to $10$ for the WMAP 5 year data at low resolution (i.e. nside = $16$). 
Blue slices are for the not iterated case and the red ones for the iterated case.
The blue plus represent the not iterated {\it BolPol} estimate with error bars (blue horizontal line) and the red plus the iterated ones.}

\end{figure*}



\begin{figure*}
\includegraphics[width=18cm,height=15cm,angle=0]{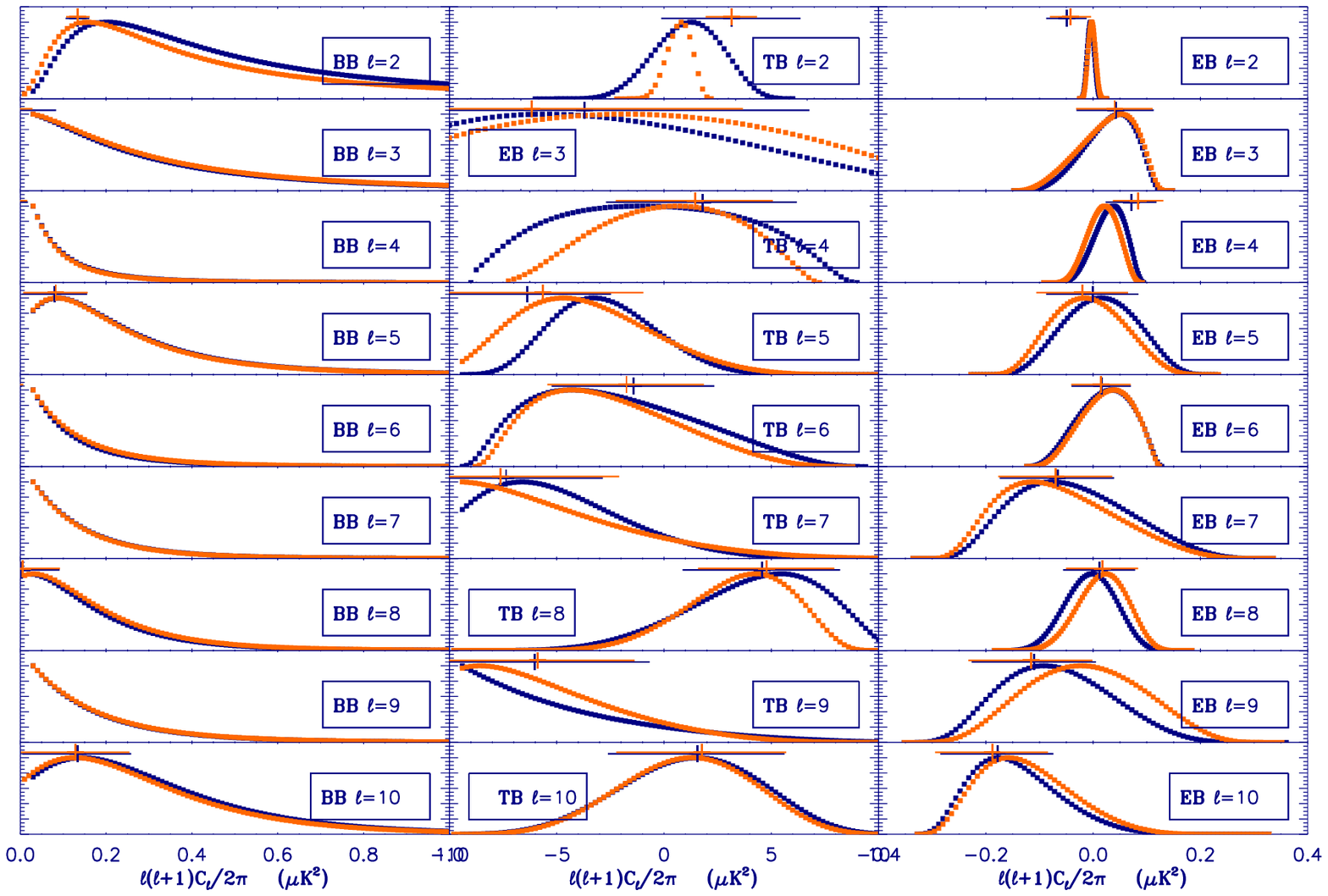}

\label{slices3}
\caption{Likelihood Slices for BB, TB and EB from $\ell=2$ to $10$ for the WMAP 5 year data at low resolution (i.e. nside = $16$). 
Blue slices are for the not iterated case and the red ones for the iterated case.
The blue plus represent the not iterated {\it BolPol} estimate with error bars (blue horizontal line) and 
the red plus the iterated ones.}
\end{figure*}




We present the results obtained by our implementation of the QML estimator on the low resolution WMAP5 maps described in the previous section
and those obtained by the WMAP team in Figs.~1-4,
respectively for TT, EE, TE and BB spectra. 
In the top panels we show the 
estimates of {\it BolPol} with error bars (dark blue or red, see below) and the (pseudo-$C_\ell$)  
estimates obtained by the WMAP team with error bars (light blue). The {\it BolPol} estimates in dark blue are obtained 
by using as fiducial spectrum the theoretical WMAP5 best-fit \citep{dunkley_wmap5}, a $\tau \Lambda$CDM cosmological model with 
$\Omega_b h^2 = 0.0227$, $\Omega_c h^2 = 0.108$, $H_0=72.4 km s^{-1} Mpc^{-1}$, $\tau = 0.089$, $n_s=0.961$, 
$A_s=2.41 \times 10^{-9}$ (at $k=0.002 Mpc^{-1}$). Error bars loose dependence on the fiducial model by iterating the QML: we use 
the first run of {\it BolPol} 
\footnote{As fiducial spectra for the iterated {\it BolPol} run we use the first $TT$ and $TE$ {\it BolPol} estimates 
and leave the $EE$ as given in the previous fiducial model when possible. The fiducial spectra of BB, TB, EB for the iterated {\it BolPol} run 
are set to zero.} 
to obtain another set of estimates with relative error bar, which is plotted in red. The iterated estimates are always 
very close to those obtained with the WMAP5 best-fit as fiducial model: this means that the QML estimated are sufficiently stable with respect to 
iteration. The same does not happen to the error bars: in particular the error on the TT quadrupole is substantially smaller, and 
a decrease of the error also occurs for $\ell=3 \,, 4$ in temperature. Our estimate for the octupole in BB is consistent with zero and 
very different from the one obtained by WMAP (note however that the WMAP likelihood slice for $C^{BB}_{\ell=3}$ is not anomalous as the 
WMAP pseudo-$C_\ell$ estimate \citep{nolta_wmap5}).

In the lower panels of Figs.~1-4
we show the differences between the 
sets of estimates in unit of sigma (same conventions as upper panels). 
More precisely we show $(C_{\ell}^{BolPol}-C_{\ell}^{WMAP})/ \sigma$ where the {\it BolPol} estimates and $\sigma$
are given by:  {\it BolPol} estimates with {\it BolPol} error bars for the no-iteration case (solid dark blue lines); {\it BolPol} estimates in the no-iteration case and WMAP error bars (dashed light blue line); {\it BolPol} estimates with {\it BolPol} error bars for the iteration case (solid red line).
These lower plots show and quantify the differences among the considered sets of estimates. 
Our estimate of the TT spectrum is within 1$\sigma$ from the one computed by the WMAP team.
For the other spectra we find larger differences (even 2$\sigma$). The current joint analysis is very different from the one
performed by the WMAP team and it is obvious to expect larger differences for the more delicate spectra (where the Signal to Noise ratio is lower).

We list now the reduced $\chi^2$ values for the iterated {\it BolPol} estimates from $\ell=2$ to $\ell=32$
with the fiducial input model:
$\chi^2_{TT} = 4.423$ but excluding the quadrupole this value decreases to $1.079$
recovering so the anomaly of  the low quadrupole value of TT; for the other reduced $\chi^2$-values we find
$\chi^2_{TE} = 0.785$, $\chi^2_{EE} = 1.422$, $\chi^2_{BB} = 1.607$. 

In Fig. 5, we plot the first and iterated {\it BolPol} estimates and relative error bars (in blue and red, respectively) 
for TB and EB. 
We do not plot the estimates by the WMAP team since these are not provided in the LAMBDA site. 
Also for these parity-odd correlators the QML estimates are very stable with respect to the iteration.
Note how the error bars in TB change 
(due to substantially different fiducial model in TT in the iterated run), whereas those in EB do not (since the fiducial EE and BB 
spectra are mainly unchanged during the iteration). The TB null reduced $\chi^2$ for $\ell=2-23$ is $1.34$ 
to be compared with $0.97$ quoted for $\ell=24-450$ by the WMAP team. The EB null reduced $\chi^2$ for $\ell=2-23$ is $1.14$. 

At low multipoles, symmetric error bars, such as those provided by the Fisher matrix, are just an approximation, since we know 
that the likelihood for $C_\ell$ is far from a symmetric Gaussian.
We therefore evaluate the conditional likelihood slices for the six spectra from $\ell=2$ to $10$; we present these results in 
Figs. 6-7. As for the QML, we compute the slices on the WMAP5 best-fit (blue points) and on the {\it BolPol} estimates (orange points obtained 
with the same fiducial used for the 
iterated QML run). It is important to note that the peaks of the likelihood slices are very different for the two sets 
of conditionings: the quadrupole in temperature is the most striking example. We do not observe such dependence on the fiducial model 
in the estimates in the QML method. As for the errors of the QML method, also the shape of the conditional slices depend on the fiducial 
model. The trend in the iteration is the same of the one observed for the QML: confidence levels shrink by inserting the iterated QML estimate as 
conditioning.
In Figs.~6-7 we also plot the QML estimate with error bars as reference (blue plus for the first {\it BolPol} run and orange for the iterated one). 
The peaks of the likelihood slices are always consistent with the
QML estimates within the error bars for both conditioning: however, an excellent agreement 
emerges when the QML estimates are used as conditioning.

Few results deserve to be commented for their cosmological importance.
Our estimate (position of the peak) for the $\ell (\ell+1) C_{\ell}^{TT}/(2 \pi)|_{\ell=2}$ is $165 \, \mu K^2$ for the iterated QML (pixel base likelihood code). 
Basing on our pixel likelihood code, the conditional likelihood (normalized to 1) 
for $\ell (\ell+1) C_{\ell}^{TT}/(2 \pi)|_{\ell=2}$ is larger than $0.05$ between $50 \, \mu K^2$ and $1305 \, \mu K^2$. 
This last value has to be compared with the one given in Fig. 1 of \cite{dunkley_wmap5},
in which we read that $\ell (\ell+1) C_{\ell}^{TT}/(2 \pi)|_{\ell=2}$ is approximatively less than $1600 \, \mu K^2$ at the same confidence level.

\begin{figure}
\includegraphics[height=5cm,angle=0]{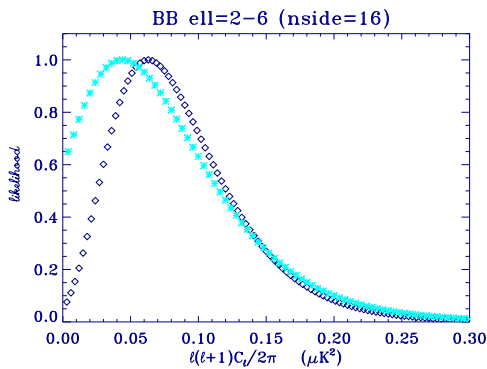}
\caption{Upper bounds on BB. $y$ axis: Likelihood slice for BB. $x$ axis $\ell (\ell+1) C_\ell^{BB}/(2 \pi)$ binning from $\ell =2 $ to $6$. 
Dark blue diamonds represent the likelihood slice obtained sampling on band powers. 
Light blue crosses represent the likelihood slice obtained sampling on power spectrum.
See also the text.}
\label{upper}
\end{figure}

We also obtain constraints on $BB$. In Fig.~\ref{upper} we show the conditional likelihood slice 
for the band powers, $\ell (\ell+1) C_\ell^{BB}/(2 \pi)$ binning from $\ell =2 $ to $6$. 
Dark blue diamonds of Fig.~\ref{upper} represent the likelihood slice we obtain
sampling on band powers. 
We find $\ell (\ell+1) \, C_{\ell}^{BB} /(2 \pi)|_{2-6}< 0.18 \mu K^2$ at $95\%$ of C.L. 
(this bound decreases to $0.13 \mu K^2$ at  $95\%$ of C.L. averaging over $\ell=2-10$).
Light blue crosses in Fig.~\ref{upper} represent the conditional likelihood slice we obtain sampling on $C_{\ell}$.
In this case we find $\ell (\ell+1) \, C_{\ell}^{BB} /(2 \pi)|_{2-6}<  0.19 \mu K^2$ at $95\%$ of C.L. 
(this bound decreases to $ 0.11 \, \mu K^2$ at  $95\%$ of C.L. averaging over $\ell=2-10$).
Some comments are in order: whereas it is expected to obtain different results sampling on different quantities, one may tentatively interpret the sampling on band powers more similar to a nearly scale-invariant spectrum of gravitational waves and the one on $C_{\ell}^{BB}$ similar to residual constant noise bias or a systematic.
Whereas we do not find evidence for a $BB$ signal one may wonder about the different shapes of the conditional likelihood slices shown in Fig.~\ref{upper}: as a matter of fact, the quadrupole is the single multipole which contributes more to this discrepancy (see also the conditional likelihood slice for $C_2^{BB}$ in Fig.~7%
. Excluding the quadrupole from the analysis, the conditional likelihood slice sampled on band powers yields an upper limit of $\ell (\ell+1) \, C_{\ell}^{BB} /(2 \pi)|_{3-6}< 0.13 \mu \, K^2$ at $95 \%$ CL. While discrepancy is not statistically significant, the fact that it is mainly due to a single multipole might be ponting at a residual contamination in the WMAP 5 years map mostly affecting the quadrupole.


\section{Discussions and Conclusions}
\label{sect:concl}

We have performed a new estimate of the CMB angular power spectra at low multipoles from low resolution maps of the five years WMAP data.

The QML estimates are found in agreement with the pseudo-$C_\ell$ WMAP ones: 
the best agreement is for the TT spectrum and differences at the level of $2-3 \, \sigma$ are found in EE and BB.

Whereas the QML is not used by the WMAP team, a pixel likelihood code is employed and results are given in 
\cite{dunkley_wmap5}, \cite{nolta_wmap5}. The difference between the conditional likelihood slices given here and those obtained by the WMAP team 
are due to our exact treatment of the covariance matrix: the approximation used by the WMAP team requires less computational resources but has the 
drawback of allowing negative values for conditional slices of EE and BB whereas ours are always positive. The origin of the difference in the constraint in 
$C_\ell^{BB}$ for $\ell=2-6$ is partially due to this difference. 

\label{sect:concl}

\section*{Acknowledgements}

We acknowledge the use of the BCX at CINECA under the agreement INAF/CINECA and the use of computing facility at NERSC.
We thank the Planck CTP working group for stimulating and fruitful interaction.
We are grateful to Joanna Dunkley and Eiichiro Komatsu for helpful discussions and clarifications.
We acknowledge use of the HEALPix \citep{gorski} software and analysis package for
deriving the results in this paper.  
We acknowledge the use of the Legacy Archive for Microwave Background Data Analysis (LAMBDA). 
Support for LAMBDA is provided by the NASA Office of Space Science.
The ASI contract Planck LFI activity of Phase E2 is acknowledged.



\begin{thebibliography}{}


\bibitem[\protect\citeauthoryear{Benabed et al.}{2009}]{teasing}
  Benabed K., Cardoso J.~F., Prunet S. and Hivon E.,
  arXiv:0901.4537 [astro-ph.CO].

\bibitem[\protect\citeauthoryear{Bond, Jaffe and Knox}{1998}]{bond}
  Bond J.~R., Jaffe A.~H. and Knox L.,
  Phys.\ Rev.\  D {\bf 57}, 2117 (1998)
  [arXiv:astro-ph/9708203].

\bibitem[\protect\citeauthoryear{Chu et al.}{2005}]{2005PhRvD..71j3002C}
Chu M., Eriksen H.~K., Knox L., G{\'o}rski K.~M., Jewell J.~B., Larson
D.~L., O'Dwyer I.~J., Wandelt B.~D., 2005, PhRvD, 71, 103002


\bibitem[\protect\citeauthoryear{Dickinson et
al.}{2007}]{2007AAS...211.9001D} 
Dickinson C., Eriksen H.~K., Jewell
J.,  Banday A.~J., Gorski K.~M., Lawrence C.~R., 2007, AAS, 38, 883


\bibitem[\protect\citeauthoryear{Dunkley et al.}{2009}]{dunkley_wmap5}
Dunkley J. et al. (WMAP), apj SS 180 (2009) 306, arXiv:0803.0586 [astro-ph].

\bibitem[\protect\citeauthoryear{Efstathiou}{2004}]{efstathiou}
Efstathiou G., Mon.\ Not.\ Roy.\ Astron.\ Soc.\  {\bf 348}, 885 (2004)

\bibitem[\protect\citeauthoryear{Efstathiou}{2004}]{efstathiou}
  Efstathiou G.,
  Mon.\ Not.\ Roy.\ Astron.\ Soc.\  {\bf 349}, 603 (2004)
  [arXiv:astro-ph/0307515].

\bibitem[\protect\citeauthoryear{Eriksen et al.}{2004}]{eriksen}
  Eriksen H.~K. {\it et al.},
  Astrophys.\ J.\ Suppl.\  {\bf 155}, 227 (2004)
  [arXiv:astro-ph/0407028].

\bibitem[\protect\citeauthoryear{Eriksen et al.}{2007}]{2007ApJ...665L...1E} 
Eriksen H.~K., Huey G., Banday A.~J.,
G{\'o}rski K.~M., Jewell J.~B., O'Dwyer I.~J., Wandelt B.~D., 2007, ApJ,
665, L1

\bibitem[\protect\citeauthoryear{Friedman et al.}{2009}]{QUAD}
  Friedman R.~B. {\it et al.}  [QUaD collaboration],
  arXiv:0901.4334 [astro-ph.CO].

\bibitem[\protect\citeauthoryear{Gorski et al.}{2005}]{gorski}
Gorski K.M., Hivon E., Banday A.J., Wandelt B.D., Hansen F.K., Reinecke M. and Bartelmann M., 2005,
HEALPix: A Framework for High-resolution Discretization and Fast Analysis of Data Distributed on the Sphere, Ap.J., 622, 759-771

\bibitem[\protect\citeauthoryear{Grain et al.}{2009}]{grain}
  Grain J., Tristram M. and Stompor R.,
  arXiv:0903.2350 [astro-ph.CO].

 \bibitem[\protect\citeauthoryear{Hauser and Peebles}{1974}]{hauser}
  Hauser, M. G.; Peebles, P. J. E., Astrophysical Journal, Vol. 185, pp. 757-786 (1973)

\bibitem[\protect\citeauthoryear{Hinshaw et al.}{2003}]{hinshaw1}
  Hinshaw G. {\it et al.}  [WMAP Collaboration],
  Astrophys.\ J.\ Suppl.\  {\bf 148}, 135 (2003)
  [arXiv:astro-ph/0302217].

\bibitem[\protect\citeauthoryear{Hinshaw et al.}{2007}]{hinshaw}
  Hinshaw G. {\it et al.}  [WMAP Collaboration],
  Astrophys.\ J.\ Suppl.\  {\bf 170}, 288 (2007)
  [arXiv:astro-ph/0603451].

\bibitem[\protect\citeauthoryear{Hivon et al}{2002}]{master}
  Hivon E. {\it et al.},
  Astrophys.\ J.\  {\bf 567}, 2 (2002)

\bibitem[\protect\citeauthoryear{Jewell et al.}{2004}]{jewell}
  Jewell J., Levin S. and Anderson C.~H.,
  Astrophys.\ J.\  {\bf 609}, 1 (2004)
  [arXiv:astro-ph/0209560].

\bibitem[\protect\citeauthoryear{Jones et al.}{2006}]{Bo}
  Jones W.~C. {\it et al.},
  Astrophys.\ J.\  {\bf 647}, 823 (2006)
  [arXiv:astro-ph/0507494].

\bibitem[\protect\citeauthoryear{Kogut et al.}{2003}]{kogut1}
  Kogut A. {\it et al.}  [WMAP Collaboration],
  Astrophys.\ J.\ Suppl.\  {\bf 148}, 161 (2003)
  [arXiv:astro-ph/0302213].

\bibitem[\protect\citeauthoryear{Kuo et al}{2004}]{ACBAR}
  Kuo C.~I. {\it et al.}  [ACBAR collaboration],
  Astrophys.\ J.\  {\bf 600}, 32 (2004)
  [arXiv:astro-ph/0212289].

\bibitem[\protect\citeauthoryear{Larson et al.}{2007}]{2007ApJ...656..653L}
Larson D.~L., Eriksen H.~K., Wandelt B.~D., G{\'o}rski K.~M., Huey G.,
Jewell J.~B., O'Dwyer I.~J., 2007, ApJ, 656, 653 \

  \bibitem[\protect\citeauthoryear{Montroy et al.}{2006}]{Bo2}
  Montroy T.~E. {\it et al.},
  Astrophys.\ J.\  {\bf 647}, 813 (2006)
  [arXiv:astro-ph/0507514].

\bibitem[\protect\citeauthoryear{Nolta et al.}{2009}]{nolta_wmap5}
Nolta J. et al. (WMAP), apj SS 180 (2009) 296, arXiv:0803.0593 [astro-ph].

\bibitem[\protect\citeauthoryear{Page et al.}{2007}]{page}
  Page L. {\it et al.}  [WMAP Collaboration],
  Astrophys.\ J.\ Suppl.\  {\bf 170}, 335 (2007)
  [arXiv:astro-ph/0603450].

\bibitem[\protect\citeauthoryear{Pearson et al.}{2003}]{CBI}
  Pearson T.~J. {\it et al.},
  Astrophys.\ J.\  {\bf 591}, 556 (2003)
  [arXiv:astro-ph/0205388].

\bibitem[\protect\citeauthoryear{Piacentini et al.}{2006}]{Bo1}
  Piacentini F. {\it et al.},
  Astrophys.\ J.\  {\bf 647}, 833 (2006)
  [arXiv:astro-ph/0507507].

\bibitem[\protect\citeauthoryear{Polenta et al.}{2005}]{polenta}
  Polenta G. {\it et al.},
  JCAP {\bf 0511}, 001 (2005)
  [arXiv:astro-ph/0402428].

\bibitem[\protect\citeauthoryear{Pryke et al.}{2008}]{QUAD2}
  Pryke C. {\it et al.}  [QUaD collaboration],
  arXiv:0805.1944 [astro-ph].

\bibitem[\protect\citeauthoryear{Rudjord et al.}{2009}]{2009ApJ...692.1669R} 
Rudjord {\O}., Groeneboom N.~E., Eriksen H.~K., Huey G., G{\'o}rski K.~M., Jewell J.~B., 2009, ApJ, 692, 1669

\bibitem[\protect\citeauthoryear{Saha et al.}{2006}]{saha}
  Saha R., Jain P. and Souradeep T.,
  Astrophys.\ J.\  {\bf 645}, L89 (2006)
  [arXiv:astro-ph/0508383].

\bibitem[\protect\citeauthoryear{Taylor et al.}{2007}]{taylor}
  Taylor J.~F., Ashdown M.~A.~J. and Hobson M.~P.,
  arXiv:0708.2989 [astro-ph].

\bibitem[\protect\citeauthoryear{Tegmark}{1997}]{tegmark_tt}
Tegmark M., Phys.\ Rev.\  D {\bf 55}, 5895 (1997)

\bibitem[\protect\citeauthoryear{Tegmark and de Oliveira-Costa}{2001}]{tegmark_pol}
Tegmark M. and de Oliveira-Costa A., Phys.\ Rev.\  D {\bf 64} (2001) 063001

\bibitem[\protect\citeauthoryear{Wandelt et al.}{2004}]{wandelt}
  Wandelt B.~D., Larson D.~L. and Lakshminarayanan A.,
  Phys.\ Rev.\  D {\bf 70}, 083511 (2004)
  [arXiv:astro-ph/0310080].

\bibitem[\protect\citeauthoryear{Wu et al.}{2007}]{MAXIPOL}
  Wu J.~H. {\it et al.},
  Astrophys.\ J.\  {\bf 665}, 55 (2007)
  [arXiv:astro-ph/0611392].

\bibitem[\protect\citeauthoryear{Wu et al.}{2008}]{QUAD1}
  Wu E.~Y. {\it et al.}  [QUaD Collaboration],
  arXiv:0811.0618 [astro-ph].


\end{thebibliography}
\end{document}